\documentclass[a4paper]{mem}
\usepackage{natbib}
\usepackage{graphicx}
\usepackage[a4paper]{hyperref}
\idline{73}{23}
\begin{document}
   \title{Search for $\alpha$ variation in UVES spectra:\\
          Analysis of C IV and Si IV doublets towards QSO\,1101-264
}

   \author{A. F. Mart\'{\i}nez Fiorenzano, G. Vladilo \and P. Bonifacio\\ 
}
   \offprints{A. F. Mart\'{\i}nez Fiorenzano}

   \institute{INAF, Osservatorio Astronomico di Trieste. Via Tiepolo 11, 34131 Trieste, Italy\\ 
   \email{fiorenza@ts.astro.it}\\ 
              }

   \abstract{   
   Motivated by previous studies of QSO spectra that reported a 
   variation of the fine structure constant $\alpha$, a search for C IV and 
   Si IV doublets was conducted in the absorption spectrum toward QSO\,1101-264,
   obtained by VLT-UVES during the Science Verification. Seven C IV and two 
   Si IV systems were identified and accurate measurements of wavelengths over 
   the redshift range $1.1862 < z < 1.8377$ were performed. After a careful 
   selection of pairs of lines, the ``Alkali Doublet" method with a derived 
   analitical expression for the error analysis was applied to compute the 
   $\alpha$ variation. The result according in magnitud order with previous
   doublets measurements, corresponds to one Si IV system: 
   $\Delta$$\alpha$/$\alpha$ $= (- 3.09 \pm 8.46) \times 10^{-5}$.

   \keywords{quasars: individual: QSO 1101-264 --
             quasars: absorption lines --
             atomic processes   
               }
   }
   \authorrunning{A. F. Mart\'{\i}nez Fiorenzano}
   \titlerunning{Search for $\alpha$ variation in UVES spectra}
   \maketitle
%

   \begin{figure}
   \centering
  \resizebox{\hsize}{!}{ \includegraphics[clip=true]{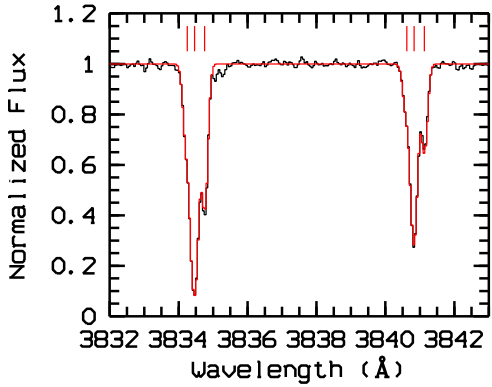}}
      \caption{Absorption of C IV at $z = 1.4767$. The leftmost line 
               in both absortions correspond to the blended and discarded 
               line. Black: observed spectrum. Red: fitted spectrum.}
         \label{figure1}
   \end{figure}

   \begin{figure}
   \centering
   \resizebox{\hsize}{!}{\includegraphics[clip=true]{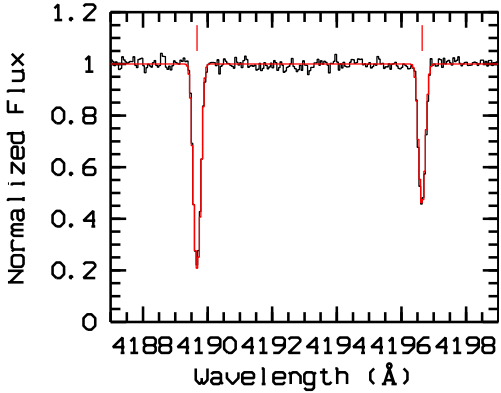}}
      \caption{Absorption of C IV at $z = 1.7061$. Black: observed spectrum. 
               Red: fitted spectrum. The sharpness of the components allows 
               us to determine the centroid accurately.}
         \label{figure2}
   \end{figure}

\section{Introduction}
   Modern theories in physics (Super Symmetry Grand Unification Theory,
   Superstring and others) predict the dependence of fundamental physical
   constants on energy, a prediction supported by high energy experiments 
   \citep{okun}, and have cosmological solutions where low-energy values   
   of these constants vary with the cosmological time 
   \citep{varshalovich}.\\
   Recent measurements of absorption lines in high resolution QSO
   spectra suggest the variation of the fine structure constant
   $\alpha$ throughout cosmological time (e.g. \citet{webb} and 
   refs. therein). The use of QSO spectra to search for an $\alpha$ variability 
   takes advantage of the many absorption lines originated in 
   clouds lying at various redshifts along the line of sight to the QSO.   
   The measurements are performed by comparing wavelength separations 
   of transitions observed at various redshifts with their corresponding 
   laboratory values at $z=0$.  Two approaches exist for this study of
   $\Delta$$\alpha$/$\alpha$: the ``Alkali Doublet" (AD) method and the
   ``Many Multiplet" (MM) method.    
   The former, first applied by \citet{bahcall}, considers
   alkaline-like ions to compare the wavelength separation of its doublets. 
   The latter, developed by \citet{dzuba99a,dzuba99b,dzuba}, 
   uses wavelengths of various transitions from different multiplets 
   and ions. In the AD method  it is common to use doublets 
   of ions like C IV and Si IV and the wavelength separation between
   $\lambda_1$ and $\lambda_2$, corresponding to the transitions
   $^2S_{1/2}\rightarrow\,^2P_{3/2}$ and $^2S_{1/2}\rightarrow\,^2P_{1/2}$
   respectively, is proportional to $\alpha^2$.\\
   From a MM method analysis based on 128 systems, \citet{webb} 
   find $\Delta$$\alpha$/$\alpha$ $= (- 0.57 \pm 0.10) \times 10^{-5}$ 
   over the redshift range $0.2 < z < 3.7$, indicating a smaller value of
   $\alpha$ in the past. 
   From an AD method analsysis of 21 Si IV doublets
   \citet{murphyb}  obtain a weighted 
   mean $\Delta$$\alpha$/$\alpha$ $= (- 0.5 \pm 1.3) \times 10^{-5}$
   at $\langle z_{abs} \rangle = 2.8$.\\        
   \citet{bahcallb}
   find $\Delta$$\alpha$/$\alpha$ $= (- 2 \pm 1.2) \times 10^{-4}$
   from an analysis of strong 
   nebular emission lines of [O III] (5007 {\AA} and 4959 {\AA}) in a QSO 
   sample over $0.16 < z < 0.80$.\\
   We have conducted a search for C IV and Si IV doublets in the absorption 
   spectrum toward QSO\,1101-264, obtained by VLT-UVES during the Science 
   Verification. Seven C IV and two Si IV systems were identified and 
   accurate measurements of wavelengths over the redshift range 
   $1.1862 < z < 1.8377$ were performed. After a careful selection of pairs 
   of lines, we applied the AD method, with an original expression 
   for the error analysis, to compute the $\alpha$ variation. 
   Here we present the results of this work. 
   \begin{figure}
   \centering
     \resizebox{\hsize}{!}{\includegraphics[clip=true]{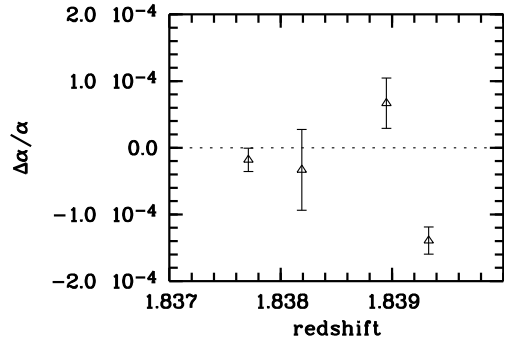}}
      \caption{Plot for the components of Si IV.}
         \label{figure3}
   \end{figure}
\noindent

\section{Measurements and Analysis}
   QSO\,1101-264 $(z_{em} = 2.145)$ was observed for 
   the Science Verification of the UVES Spectrograph at VLT.
   All the work done on the spectrum was performed under the MIDAS package.   
   Data reduction was carried out by the Science Verification Team and for 
   our purpose, the reduced spectrum was converted from air wavelengths 
   to vacuum wavelengths.\\
   A total of nine systems among the C IV (1548{\AA}, 1550{\AA}) and 
   Si IV (1393{\AA}, 1402{\AA}) ions were identified. A code for line 
   fitting considering Voigt profiles and based on the reduced $\chi^{2}$ test 
   as parameter of goodness of the fit was used. The laboratory wavelengths 
   considered in the present analysis are: C IV (1548.204{\AA}, 1550.781{\AA}) 
   and Si IV (1393.76018{\AA}, 1402.77291{\AA}) measured by 
   \citet{griesmann}, the best available up to date.\\
   The sample contains a total of 47 doublets, not all suitable for the 
   desired measure of $\Delta$$\alpha$/$\alpha$ due to complicated 
   profiles, some of them contaminated or blended. For that reason a 
   rigorous selection of lines was made, choosing the best resolved 
   profiles, outside the Lyman forest to avoid contaminations, discarding 
   asymmetric shapes and focusing on narrow absorptions because centroids are better
   determined than in broader ones. We present only those selected in 
   Table~\ref{Delta}.\\
   The synthetic spectrum for every doublet was constructed by using the 
   minimum number of components, until a very good shape reproducing the 
   observed spectrum was reached.\\
   The C IV system at $z = 1.4767$ in Figure~\ref{figure1} helps to illustrate
   the fitting procedure. After several line fitting runs with different 
   starting guess values around the minima, a poor fit was obtained. 
   To improve it, new lines were added, paying attention to the 
   slightly asymmetric shape of the spectrum toward the left. It allows to 
   determine better the wavelengths of the strong minima but the others do 
   not represent a reliable absorption to be considered.\\ 
   The C IV system at $z = 1.7061$ is a fortunate case because it is a simple 
   pair of strong lines (see Figure~\ref{figure2}). It is a very good example 
   of the available high quality instruments displaying a spectrum with 
   $R \approx 45000$ and $S/N \approx 60$. The selection of appropriate centroid 
   wavelengths was based on this profile where the line fitting provided 
   wavelengths uncertainties of 
   $\sigma_{\lambda_{z1}}$=\,$0.0015${\AA} and 
   $\sigma_{\lambda_{z2}}$=\,$0.0021${\AA}. Because, somehow, a complex velocity 
   structure is present in all systems, the most symmetric shapes were selected 
   qualitatively and the corresponding $\sigma_{\lambda_{z1}}$
   and $\sigma_{\lambda_{z2}}$ values served as quantitative criterion to 
   ensure the best wavelengths to be included in the calculations.\\
   Considering a possible small variation of $\alpha$, 
   \citet{varshalovich} use the approximate formula
   \begin{equation}
       \frac{\alpha_{z}-\alpha}{\alpha} =
         \frac{\Delta\alpha}{\alpha} = \frac{c_{r}}{2}
         \left [ \frac{\left (\Delta\lambda/\lambda \right )_{z}}
                    {\left (\Delta\lambda/\lambda \right )_{0}} - 1
         \right ] 
   \end{equation}   
   where $\lambda = \frac{1}{2} \left (\lambda_{1} + \lambda_{2} \right )$;
   $\left (\Delta\lambda/\lambda \right )_{z}$ and 
   $\left (\Delta\lambda/\lambda \right )_{0}$
   represent the doublet separation for the absorption at redshift 
   $z$ and at the laboratory, respectively; 
   $c_{r} \approx 
   \left ( \delta q_{1} - \delta q_{2} \right )/
   \left ( \delta q_{1} - 2 \delta q_{2} \right ) $ 
   is a correction term given by \citet{murphy}, with $q_{1}$ 
   and $q_{2}$ coefficients representing the relativistic correction to 
   the energy for a particular transition, calculated for many elements 
   by \citet{dzuba99b}. The correction coefficients $c_{r}$ 
   are: 1.1758 for C IV and 0.8914 for Si IV. An analitic expression for 
   the error analysis can be obtained through an aproximation for the 
   standard deviation as 
   $\Delta$$\alpha$/$\alpha$ = $f \left ( \lambda_{z1},\lambda_{z2} \right )$:
   \begin{equation}
      \sigma_{f}^{2} \approx \sigma_{\lambda_{z1}}^{2} 
       \left( \frac{\partial f}{\partial \lambda_{z1}} \right)^{2} +
       \sigma_{\lambda_{z2}}^{2} 
       \left( \frac{\partial f}{\partial \lambda_{z2}} \right)^{2} + \cdots
   \end{equation}
   which, with the derivatives of eq. (1), yields the error propagation equation 
   for the AD method. Results appear in Table~\ref{Delta}, a plot for components 
   of Si IV is shown in  Figure~\ref{figure3} and averaging the four values lead to:
   $\Delta$$\alpha$/$\alpha$ $= (- 3.09 \pm 8.46) \times 10^{-5}$, where the 
   error is the standard deviation around the mean.\\ 
         
   \begin{table}
      \caption[]{$\Delta$$\alpha$/$\alpha$ calculations with
                 their corresponding standard deviations.}
         \label{Delta}
     $$ 
         \begin{array}{ccccp{0.5\linewidth}cccc}
            \hline
            \noalign{\smallskip}
            Ion,\,z & \frac{\Delta\alpha}{\alpha} & \sigma_{\Delta\alpha/\alpha}\\
            \noalign{\smallskip}
            \hline
            \noalign{\smallskip}
            C IV,\,\,1.4767 & -1.2330\times 10^{-3} & 8.4913\times 10^{-5}\\
            C IV,\,\,1.4769 & -5.0416\times 10^{-4} & 1.9919\times 10^{-4}\\
            C IV,\,\,1.7061 & -8.5510\times 10^{-4} & 2.1750\times 10^{-4}\\
            C IV,\,\,1.8377 & -9.4692\times 10^{-4} & 1.1591\times 10^{-4}\\
            C IV,\,\,1.8385 & -4.6644\times 10^{-4} & 2.5231\times 10^{-4}\\
            C IV,\,\,1.8389 & -1.0835\times 10^{-3} & 1.0289\times 10^{-4}\\
            Si IV,\,\,1.8377 & -1.8157\times 10^{-5} & 1.7412\times 10^{-5}\\
            Si IV,\,\,1.8381 & -3.3109\times 10^{-5} & 6.0675\times 10^{-5}\\  
            Si IV,\,\,1.8389 &  6.6867\times 10^{-5} & 3.7639\times 10^{-5}\\                    
            Si IV,\,\,1.8393 & -1.3921\times 10^{-4} & 2.0282\times 10^{-5}\\                    
            \noalign{\smallskip}
            \hline
         \end{array}
     $$ 

   \end{table}
\section{Conclusions}
   For the first time, UVES data have been employed to compute
   $\Delta$$\alpha$/$\alpha$ with the AD method.
   The spectrum under study has a very good quality, comparable, 
   or better, than spectra used in previous work. 
   Despite the stringent selection criteria applied to find suitable 
   absorption profiles for measuring central wavelengths, an acceptable 
   number of doublets was obtained to carry out the desired calculations.\\
   Even if the absorption profiles of C IV have very high quality, the   
   measurements errors obtained from this ion are bigger by one order of 
   magnitude than those obtained from Si IV. In addition, the   
   $\Delta$$\alpha$/$\alpha$ values provided by the C\,IV doublets
   disagree in order of magnitude with those provided by Si IV, that 
   appear to be consistent with previous determinations 
   \citep{murphyb}. These differences between C IV and Si IV 
   data are probably due to the better determination of laboratory 
   wavelengths of Si IV in comparison to C IV wavelengths. Until better 
   laboratory data for C IV is available, we consider only the result 
   obtained from the components of the  Si IV system at $z \simeq 1.84$. 
   The resulting 
   $\Delta$$\alpha$/$\alpha$ $= (- 3.09 \pm 8.46) \times 10^{-5}$
   does not support a change of $\alpha$ at such redshift. However, despite 
   the result consistent with zero, our $\Delta$$\alpha$/$\alpha$ values show 
   a negative sign, which is consistent with the findings of all previous 
   determinations of $\Delta$$\alpha$/$\alpha$ 
   \citep{varshalovich,webb,murphyb,bahcallb}. 
   This remarkably consistent indication for a possible variation of $\alpha$
   certainly deserves further investigation on a large number of systems,
   aimed at reducing the final error bar.  
   The selection process of the best absorption lines, made to improve the    
   $\Delta$$\alpha$/$\alpha$ determination, reduces the quantity of doublet 
   systems available in a single QSO spectrum in a sensitive way. This is a 
   motivation to study many more QSO's spectra to enhance the data sample, 
   concentrating the analysis on the Si IV transitions, which have well 
   determined laboratory data.

\begin{acknowledgements}
   AFMF thanks Valentina D'Odorico, Miriam Centuri\'{o}n and Paolo Molaro 
   for help and advice, the librarians and the people of the Osservatorio 
   Astronomico di Trieste. This work was supported by a scholarship from 
   the Ministero degli Affari Esteri, with the support of the Ambasciata 
   d'Italia in Colombia and Istituto italiano di Cultura in Bogot\'{a}, 
   Colombia.
\end{acknowledgements}

\bibliographystyle{aa}

\end{document}